# Electronic bandstructure and van der Waals coupling of ReSe$_2$ revealed by high-resolution angle-resolved photoemission spectroscopy

Lewis S Hart[1], James L Webb[1], Sara Dale[1], Simon J. Bending[1], Marcin Mucha-Kruczynski[1], Daniel Wolverson[1*], Chaoyu Chen[2], José Avila[2], Maria C. Asensio[2*]

ReSe$_2$ and ReS$_2$ are unusual compounds amongst the layered transition metal dichalcogenides as a result of their low symmetry, with a characteristic in-plane anisotropy due to in-plane rhenium 'chains'. They preserve inversion symmetry independent of the number of layers and, in contrast to more well-known transition metal dichalcogenides, bulk and few-monolayer Re-TMD compounds have been proposed to behave as electronically and vibrational decoupled layers. Here, we probe for the first time the electronic band structure of bulk ReSe$_2$ by direct nanoscale angle-resolved photoemission spectroscopy. We find a highly anisotropic in- and out-of-plane electronic structure, with the valence band maxima located away from any particular high-symmetry direction. The effective mass doubles its value perpendicular to the Re chains and the interlayer van der Waals coupling generates significant electronic dispersion normal to the layers. Our density functional theory calculations, including spin-orbit effects, are in excellent agreement with these experimental findings.

[1]Centre for Nanoscience and Nanotechnology, Department of Physics, University of Bath, Bath BA2 7AY, United Kingdom
[2]Synchrotron SOLEIL, Saint Aubin, and Université Paris-Saclay, BP 48 91192 Gif-sur-Yvette, France

Correspondence should be addressed to D. Wolverson (d.wolverson@bath.ac.uk) or M. C. Asensio (maria-carmen.asensio@synchrotron-soleil.fr)

The layered TMD family includes a rich palette of superconductors, metals[1] and semiconductors with direct and indirect gaps, and offers fascinating possibilities for the realisation of nanoscale electronic, optoelectronic and photonic devices through the assembly of heterostructures[2]. These may include dissimilar TMDs, but also graphene and boron nitride[3, 4]. Typical semiconducting TMDs ($MoS_2$, $WS_2$, $WSe_2$) are hexagonal, with inversion symmetry in the bulk which is absent for the monolayer. This symmetry-breaking in monolayers leads to a finite SO splitting and to the non-equivalence of the $K^+$ and $K^-$ valleys[5]. The exciton binding energies and SO splittings are typically large[5, 6], giving optical access to well-defined spin-valley states even at room temperature. At the same time, the direct gap of monolayers appears at the K points, so that circularly polarised excitation can address selectively either $K^+$ or $K^-$ valleys. By alloying, for example, $MoSe_2$ with $WSe_2$, the magnitude of the SO splitting may be varied, and this allows tuning of the above effects[7].

However, the TMD family also includes materials which do not conform to the typical expectations above[1], and this much less well-known group of TMDs expands the range of possible heterostructures. One such material is $ReSe_2$ (and the closely-related $ReS_2$) in which the only symmetry operation is inversion[8, 9, 10, 11]. In contrast to typical TMDs, an inversion centre in Re-TMDs is preserved *even in monolayers*, so that few-layer Re-TMDs are expected to have zero SO splitting independent of layer number[12]. Nevertheless, spin-orbit effects still modify the band structure of $ReSe_2$, shifting the transition metal (Re) *d*-orbitals that make up its band edges[13]. Consequently, perturbations that break inversion symmetry, such as alloying[14] or external electric fields[15], may allow one to manipulate the valence band SO splitting in $ReSe_2$ or $ReS_2$, and this splitting will grow *from zero* rapidly on applying a given perturbation.

Thus, the Re-TMDs promise a new means to control SO effects in few-layer semiconductor heterostructures. Being highly anisotropic 2D materials, they also offer new possibilities as hyperbolic plasmonic materials[16] or polarisation-sensitive photodetectors[17, 18, 19, 20]. Interest in anisotropic 2D materials is growing rapidly, with the isolation of few-layer black phosphorus and analogues such as GeS; relative to these materials, however, we find $ReS_2$ and $ReSe_2$ are more promising because they are stable in air[21]. In particular, the van der Waals coupling between layers has been estimated as very weak and consequently quasi-monolayer behaviour in bulk Re-TDMs has been reported based on recent micro-Raman and photoluminescence results [21]. However, recent angle-resolved Raman studies conclude that exciton-phonon coupling and more exotic interactions can be present in Re-TMD compounds[22]. Nevertheless, to predict what might be achieved using the Re-TMDs, a precise understanding of their band structure is essential. There have only been a few attempts to model either bulk or monolayer Re-TMDs[12, 23, 24] and calculations have not explored the full Brillouin zone. Furthermore, no direct band structure determination has been reported to date, though indirect data on optical absorption[25, 26, 27, 28] and transport properties[29, 30, 31, 32, 33, 34] are available. The present work addresses this lack of information for the case of bulk $ReSe_2$.

We present here the first measurements of the valence band structure of bulk $ReSe_2$, using angle-resolved photoemission (ARPES) with nanoscale spatial resolution (nano-ARPES). Our results are modelled via density functional theory (DFT) including spin-orbit (SO) effects. We find a remarkable band structure, with two valence band maxima within the first Brillouin zone and related by inversion symmetry, but not located on any special high-symmetry points or paths and, therefore, predicted to be subject to strong SO splitting under external perturbation. The ARPES data reflect the strong in-plane asymmetry peculiar to the transition metal dichalcogenides (TMDs) based on Re[8, 9, 10, 11], with very different valence band dispersions parallel or perpendicular to the Re chains that run along the crystallographic *a* direction (see Figure 1a and b). As a result, the effective mass along the Re chains is almost twice that orthogonal to them. Even more interestingly, the excitation energy dependence of the nano-ARPES data shows a striking out-of-plane dispersion, indicating that the interlayer van der Waals coupling in ReSe$_2$ is appreciable and therefore, the electronic properties on monolayer ReSe$_2$ compounds could be dramatically different to the bulk material. Finally, even if a full understanding of the momentum-resolved electronic structure of ReSe$_2$ is particularly complex due to its triclinic crystal structure, the two-fold theoretical and experimental approach taken here allows us to identify the electronic hallmark of this compound as well as how the bulk band structure relates to that of Re-TMD monolayers.

**Results**

**ARPES Fermi surface and constant energy mapping.** One of the most illuminating modes of angle-resolved photoemission (ARPES) measurement nowadays is to monitor with high energy and angular resolution the photoemission signal from states in a given energy window near the Fermi level as a function of electron wavevector parallel to the crystal surface, since it is this wavevector component that is conserved in photoemission[35]. For a three-dimensional material, this yields a section through reciprocal space, which is nearly planar, that is, having a nearly constant wavevector component normal to the sample surface. However, the deviation of this section from planarity can become important, as will be discussed below.

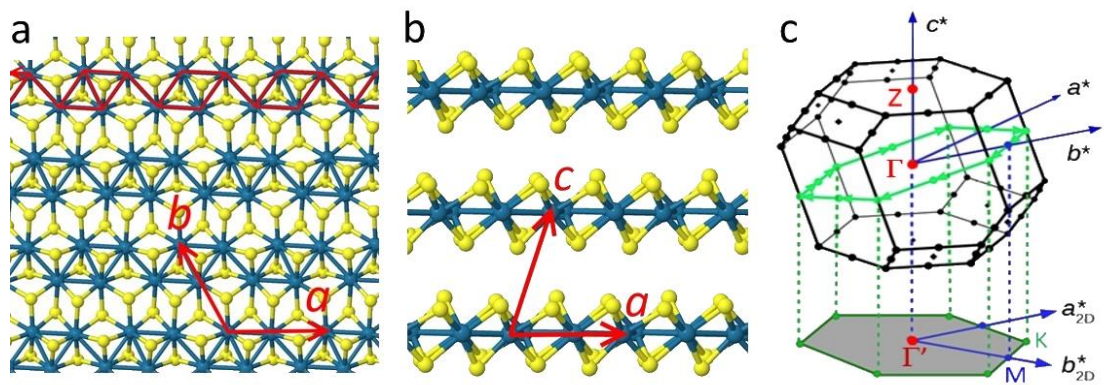

**Figure 1 Crystal structure and first Brillouin zone of triclinic ReSe$_2$.** (**a**) View of a single layer seen from above and (**b**) from the side. Re atoms are coloured blue and Se atoms are yellow. The directions of the lattice vectors *a*, *b* and *c* are indicated; *a* is defined here as the direction of the rhenium chains, highlighted in red in (**a**). (**c**) First Brillouin zone of ReSe$_2$ indicating the reciprocal lattice vectors ***a\****, ***b\*** and ***c\*** and the conventional points Γ (0, 0, 0) and Z (0, 0, ±½). The tilted green hexagon indicates a path in reciprocal space around points of the type (±½, 0, 0) pseudo-Brillouin zone defined by the projections of these points onto the real space layer plane, with basis vectors labelled $\boldsymbol{a}^*_{2D}$ and $\boldsymbol{b}^*_{2D}$ and centre Γ'.

To interpret our nano-ARPES data for bulk ReSe$_2$, we first need to discuss its crystal structure and reciprocal lattice. Figures 1a and 1b show how the Re atoms within a layer form groups of four in linked chains of rhombuses, driven by a distortion of the unstable metallic hexagonal structure into the semiconducting 1T' phase[23, 36]. These layers stack along the crystallographic *c* axis which, note, is *not* normal to the layers (Fig. 1b). Figure 1c shows the resulting Brillouin zone for the bulk material; the reciprocal lattice vectors ***a\*** and ***b\*** do not lie in the real space layer plane so that, unlike the cases of hexagonal MoS$_2$ or WS$_2$, the plane probed in ARPES does not contain these basis vectors. Instead, ARPES will (to a first approximation) map a plane normal to **c*** with a 2D quasi-unit cell consisting of the shaded irregular hexagon which is defined by the projections of **a*** and **b***. Figure 1c shows that this hexagon is a projection of the tilted hexagon that contains some of the special points of the Brillouin zone (BZ). The 2D quasi-Brillouin zone itself does not contain any special points and, consequently, the conventional labels for the high-symmetry points of a regular hexagon (K and M) are not strictly applicable. For convenience, however, we will keep these labels and number these points $K_1..K_3$ and $M_1..M_3$ later, when it is necessary to distinguish between the non-equivalent K and M directions. Likewise, the centre of the 2D quasi-unit cell, $\Gamma'$, is a point on the line joining $\Gamma$ and $Z$ points (Figure 1c), and is not necessarily the true 3D Brillouin zone centre. For a monolayer, the gray hexagon becomes the true 2D reciprocal space unit cell, so that labels $\Gamma$, K and M become strictly correct[12].

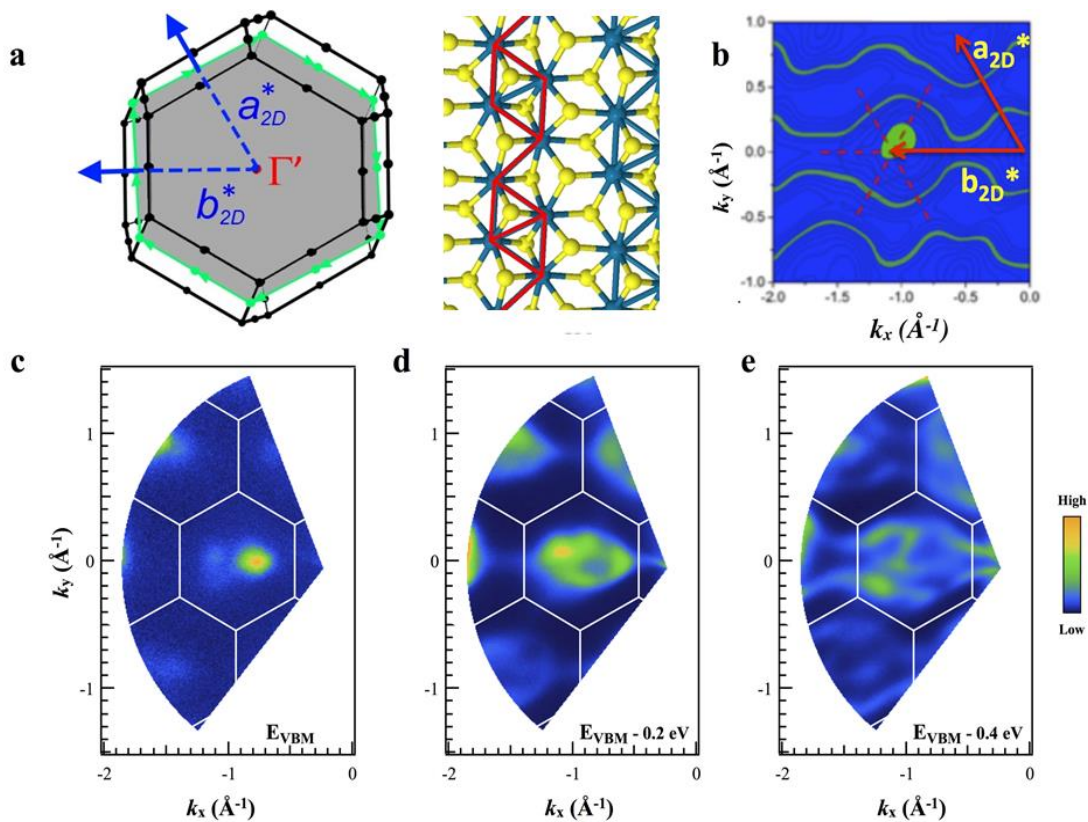

**Figure 2**. **ARPES maps of photoemission intensity as a function of in-plane wavevector** (a) View along the **c*** axis of the first Brillouin zone, showing the irregular hexagon (gray) defined by $\boldsymbol{a}^*_{2D}$ and $\boldsymbol{b}^*_{2D}$, and top view of a single ReSe$_2$ layer indicating the orientation of the chains with respect to the Brillouin zone. (b) a theoretical contour plot at 0.4 eV below the valence band maximum. The dotted cross indicates the position of a $\Gamma'$ point. For comparison, in Figure S1 of the supplementary information all experimental contour plots are compared with the corresponding theoretical ones. (c)-(e) Experimental maps (c) near the valence band maximum (VBM); (d) 0.2 eV below the VBM and (e) 0.4 eV below the VBM. In each case, the dotted cross indicates the position of a $\Gamma'$ point.

We now turn to the nano-ARPES maps shown in Figure 2. The 2D images show photoemission intensity as a function of in-plane wavevector ($k_x$, $k_y$) for states at a series of three energies near and just below the valence band maximum (VBM), using a fixed excitation photon energy of 100 eV; we discuss the implications of the choice of excitation energy below. Using a gold sample as reference *in situ*, the energy difference between the Fermi level and the valence band maximum located in Figure 2c has been precisely determined to be 1.4 ± 0.025 eV, close to the direct excitonic optical band gaps of ReSe$_2$ at low temperatures (1.386 and 1.409 eV[37, 38] at 15 K). This is consistent with the fact that the present ReSe$_2$ samples are highly *n*-type as indicated by the transport characteristics of FET structures made from the same batch of material (see Supplementary Information Figs. S2 and S3) so that the Fermi level is close to the conduction band. More usually, ReSe$_2$ is found to be p-type[31, 39] though this is not universally the case[40]. Given this information, we can label the constant energy surfaces of Fig 2c-e with the binding energies (1.4 eV, 1.6 eV and 1.8eV) or (0 eV, 0.2 eV and 0.4eV), depending on whether the Fermi energy or the valence band maximum respectively is taken as a reference (see Fig. 2).

The 2D quasi-Brillouin zone (Figs 1c and 2a) is clear in the distribution of the maxima in Figure 2(c-e) and is in excellent agreement with lattice vectors calculated from the crystal axes determined by X-ray diffraction[8]. Even more striking is the anisotropy these maps show between $k_x$ and $k_y$ directions. This is seen most clearly in Figure 2(e) where the contours of photoemission intensity form wavy 'ribbons' running along the $k_x$ direction parallel to $\boldsymbol{b}^*_{2D}$; this is the direction perpendicular to the Re chains, which we define to be along the real space vector $\boldsymbol{a}$ (Figure 1). This result reveals a much flatter valence band dispersion for carriers moving perpendicular to the chains compared to those moving along the chains (we return to this point below). For comparison, Figure S1 of the supplementary information shows the contours of constant energy calculated via DFT for a section through the Brillouin zone, probing the uppermost VB state. The energies of the contours are the same energies at which the experimental contours have been measured.

The experimental and calculated constant energy maps throughout the reciprocal space unit cell reflect the characteristic signatures of the density of states at a given energy and momentum. Note that the calculated images are not simulations of the ARPES signals, as the latter depend also on the photoemission matrix elements. Nevertheless, both sets of patterns show a remarkable agreement. Firstly, it is clearly noticeable that the highest–energy states do not appear centred on a Γ' point, but are displaced to either side. Secondly, for photoelectrons of higher binding energies, both the nano-ARPES and the theoretical results show the development of the one-dimensional ('wavy') structure over exactly the same energy and momentum ranges (see Fig. 2b and 2e and Fig. S1 of the supplementary information). Finally, we note that this good agreement between experimental and theoretical results extends to lower-energy VB states, not just those of the uppermost band; in Figure 2b, for example, we have to include a contribution from the next band down in energy, which appears in the experimental energy range.

The in-plane anisotropy of the electronic structure of ReSe$_2$ can be investigated more deeply by recording high energy- and angular-resolution photoemission scans along selected high symmetry directions. Figure 3 shows such scans through the 2D quasi-Brillouin zone for a plane passing through the Z point in reciprocal space (see Fig 1c). Fig. 3a shows the Fermi surface map recorded in this plane, and also shows how we define the labels for the non-equivalent points $K_1..K_3$ and $M_1..M_3$. Here, we use a prime (e.g, M') to represent the projection through Z of a given point (e.g, M) to its equivalent in the next Brillouin zone. In Fig 3b, the nano-ARPES scans are shown for the six directions of type M'-Z-M and K'-Z-K'. Fig 3b shows the effects of anisotropy in several distinct ways.

Firstly, it shows unmistakably the inequivalent character of the $K_2$–Z and $M_1$–Z directions which are parallel and perpendicular respectively to the Re chains. Secondly, the dispersions are similar in the $K_1$–Z and $K_3$–Z directions, which make approximately the same angle to $K_2$ (and thus to the Re chains), and the dispersions in the $M_2$–Z and $M_3$–Z directions are similar to each other for the same reason. Finally, there is a clear asymmetry between the dispersions along M'-Z and M-Z, that is, either side of the Z point in the same direction. This is due to the shape of the three-dimensional Brillouin zone shown in Fig. 1c; there are no true reciprocal lattice vectors in the plane being probed, so M' is not equivalent to M. The fact that the Brillouin zone has to be considered as three-dimensional indicates the existence of significant inter-layer coupling (we return to this point later).

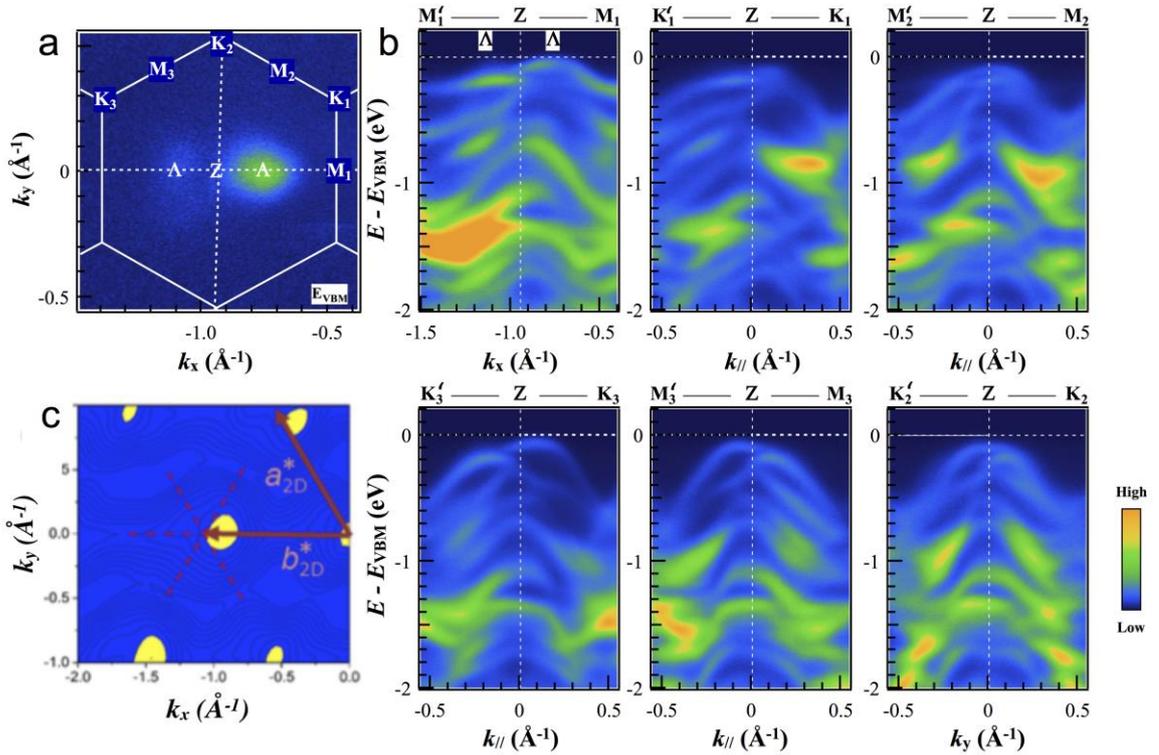

**Figure 3. Electronic band structure of ReSe$_2$ (a) ARPES constant energy plots measured with 100 eV photon energy throughout the reciprocal unit ReSe$_2$ cell.** Experimental photoemission signal as a function of in-plane momentum at an energy close to the VBM for a section through the Brillouin zone passing through the Z point, showing the special points of the quasi-Brillouin zone including the position of the local VBM (labelled Λ) **(b)** ARPES data along the M'ZM and K'ZK directions in the reciprocal space. **(c)** DFT calculations of the density of states at the top of the valence band in the plane containing the Z point, in good agreement with experimental results shown in panel **(a)**

To test the agreement of our experimental and theoretical results more closely, Figure 4 shows the measured and calculated band dispersions along the key orthogonal directions (a) Z-$K_2$ and (b) Z-$M_1$ over a large energy range (to binding energies of more than 3 eV below the Fermi level). For clarity, the experimental dispersions are second derivatives of the raw photoemission data where the colour scale represents signal strength whilst, in the calculated dispersions, the colour scale represents the projection of the VB states onto the Re $d$ orbitals, for which a broadening in energy of $\Delta E$ = 68 meV and an energy grid step of 20 meV were used. In the theoretical curves of Fig. 4a,b the occasional periodic structure is an artefact of these grid choices. The non-conservation of the initial electron momentum expressed by Eq. 1 (see below) has been taken into account in the simulated dispersions using an inner potential of 19.1 eV. The number and structure of bands within the ~2 eV energy range of this data is clear in Fig. 4a,b, and the asymmetry of the bands is once more very striking. We conclude that the present level of DFT approximation (see Methods) is adequate to describe the band structure well, but that account must be taken of the three-dimensional nature of the band structure that gives rise to the structure of Fig. 4a.

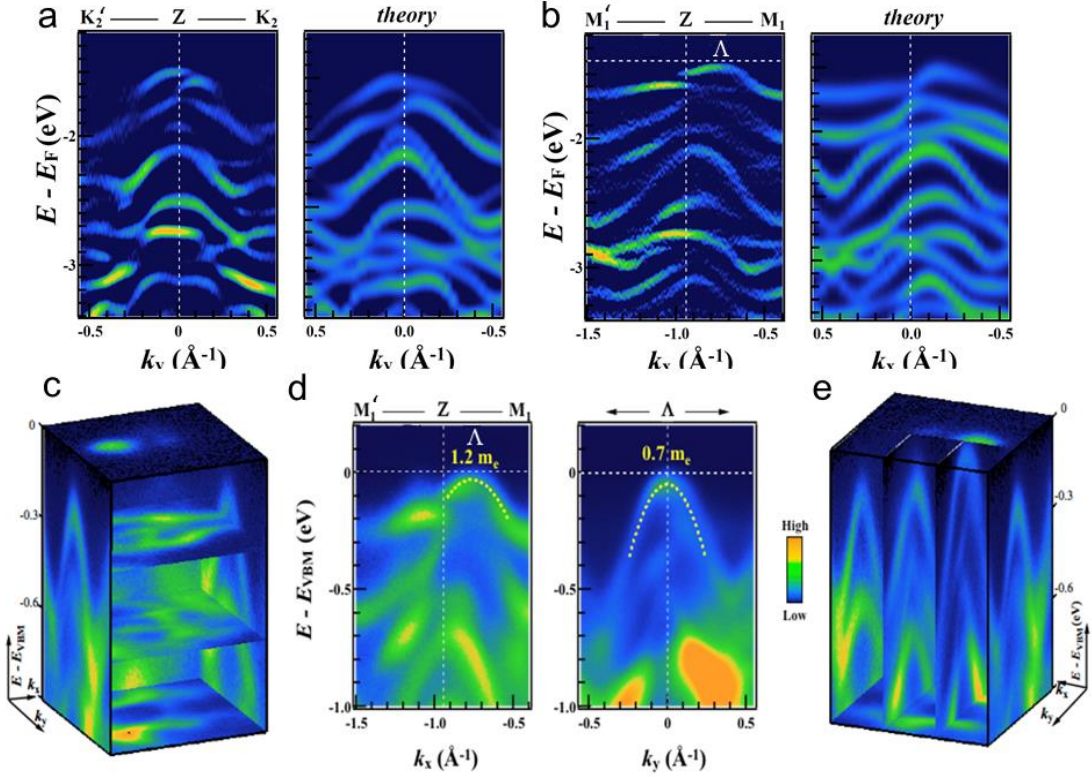

**Figure 4. Valence band dispersion.** (**a**) Measured and calculated dispersions along the $b_{2D}^*$ direction and (**b**) normal to it. In (a) and (b), the colour scale of the calculated data is arbitrary but in the same sense as the experimental data and indicates the projection of the VB states onto the Re $d$ orbitals. (c) and (e) panels show the same data of panels (a) and (b) in 3D plots. (**d**) Dispersion measured by ARPES along the $b_{2D}^*$ direction (left) and normal to it (right) passing through the point $\Lambda$. Fitted dispersions are shown as dashed lines, giving the effective masses at $\Lambda$ in these two directions.

The comprehensive ARPES datasets represented in Fig 4c,e that give the sections through the Brillouin zone of Fig 4a,b also allow us to obtain the electronic dispersions exactly at the local valence band maximum (labelled as the point $\Lambda$) both along the 1D Re chains and orthogonal to them. Measured dispersions passing through $\Lambda$ are shown in Fig. 4(d) along the $b_{2D}^*$ direction (left) towards $M_1$, and normal to $b_{2D}^*$ (right). Fitted parabolas (Fig. 4d, dotted white lines) allow a precise estimation of the degree of in-plane anisotropy. Effective masses of 0.4 $m_e$ and 1.2 $m_e$ have been directly determined along the direction of the Re-atomic chains and orthogonal to them, respectively. Clearly, the effective mass is lowest in the direction normal to $b_{2D}^*$ or, equivalently, the direction parallel to the Re chains in real space. In a 2D TMD, the phonon-limited mobility depends on the inverse square of the effective mass[34, 41, 42] so that we expect a higher mobility along the Re chains. The ARPES results thus provide a direct experimental explanation for the higher mobility in the Re chain direction found very recently for top-gate field effect transistors (FETs) based on few-layer ReSe$_2$[40] though it was noted in that work that the substrate affects the measured mobility.

The simulations showed in Figures 2, 3 and 4 account for the non-conservation of the initial electron momentum $k_z$ perpendicular to the crystal surface in photoemission[35]. This is particularly important here, since there is significant dispersion of the energy bands in the $k_z$ (*c**) direction, as shown by the calculated constant-energy surfaces plotted in Figure 5a. We proceed as follows: assuming the final state is a free electron with kinetic energy $E_{kin}$ and a parabolic dispersion starting at the inner potential $V_0$, then $k_z$ in the initial state is

$$|k_z| = \frac{\sqrt{2m}}{\hbar}\sqrt{(E_{kin}\cos^2\theta) + V_0} \qquad (1)$$

where $m$ is the free electron mass, $\theta$ is the polar emission angle, and $E_{kin} = E_{photon} - \phi - E_B$ where $E_{photon}$ is the excitation photon energy, $\phi$ is the work function of the material and $E_B$ is the binding energy of the initial state[43]. For ReSe$_2$, the inner potential $V_0$ is unknown and so its value was estimated as part of this work.

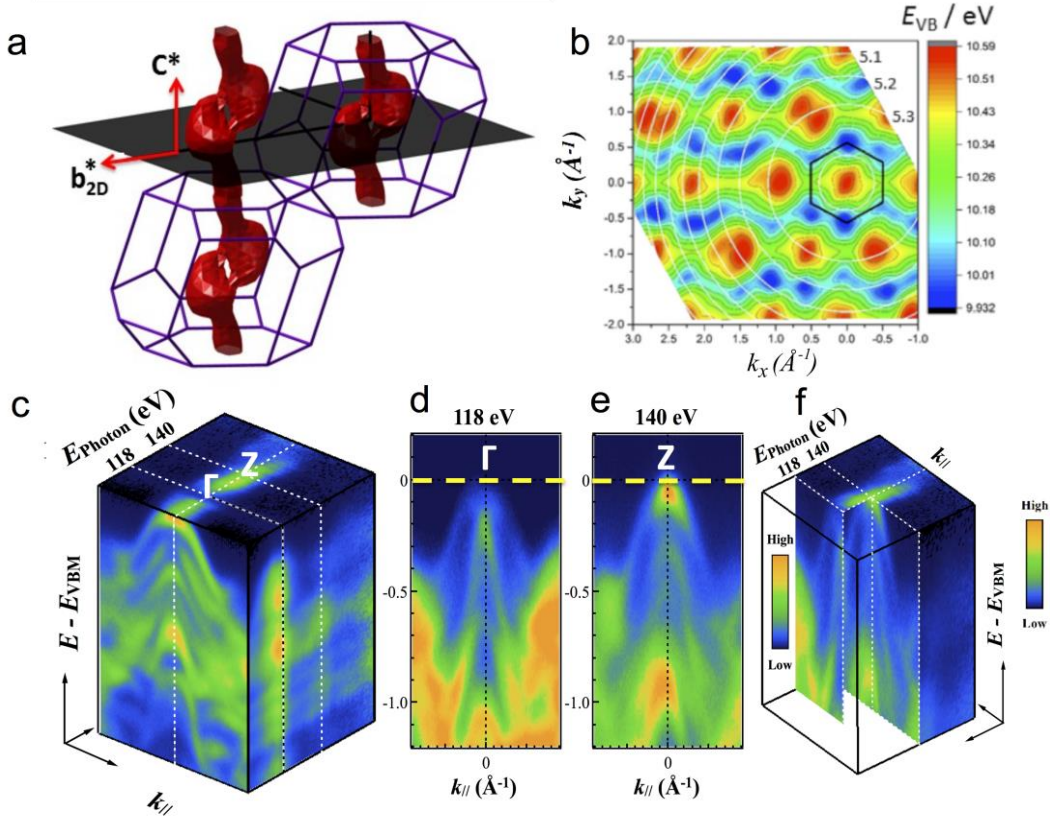

**Figure 5. Three-dimensional electronic band structure of the ReSe$_2$** **(a)** Perspective view of the Brillouin zone: red surface: the constant energy surface for the valence band states at 0.2 eV below the valence band maximum. The shaded plane shows a plane parallel to the crystal layers; this is the plane sampled in an ARPES experiment at a given excitation phonon energy (neglecting the curvature due to momentum conservation discussed in the text). **(b)** Energy $E_{VB}$ of the highest-lying valence band state as a function of in-plane momenta $k_x$ and $k_y$, calculated taking into account the variation of the momentum $k_z$ normal to the layer expressed by equation 1. White circles indicate contours of constant $k_z$ (three are labelled, top right, by their values in units of $2\pi/c^*$) and the black hexagon shows the in-plane quasi-

unit cell of Figure 1. **(c)** and **(f)** nano-ARPES signal (blue=low to orange=high) as a function of energy below the Fermi energy (vertical axis) and in-plane momentum $k_{xy}$, for excitation energies of 118 and 140 eV (left and right respectively). **(d)** and **(e)** panels show the nano-ARPES electronic dispersion of the valence bands at the Γ and Z points of the 3D Brillouin unit cell.

To determine $V_0$, the excitation photon energy has been varied whilst monitoring the photoemission perpendicular to the crystal surface ($\theta = 0$); we look for those photon energies for which $k_z$ in the above equation is an integer or half-integer multiple of the reciprocal lattice vector $c^*$ (the Γ and Z points respectively) Experimental data are shown in Figure 5c. The ARPES dispersion as a function of the incident photon energy for 95 eV < $E_{photon}$ < 180 eV shows clear photoemission minima and maxima at 118 and 141 eV respectively, (Fig. 5c-f and Fig. S4 of the Supplementary information). This, together with the magnitude of $c^*$ (0.984 Å$^{-1}$) gives $V_0$ = 19.1 ± 0.1 eV (see SI for further discussion of the analysis). This value is typical for similar TMDs (e.g, $V_0$ = 13 eV for WSe$_2$[44]).

For finite in-plane momentum ($\theta \neq 0$), as in the maps of Fig. 2, equation 1 shows that the value of $k_z$ in the initial state will vary as $\theta$ is varied at a given binding energy and photon energy. Therefore, the experimental section through the 3D band structure is not strictly planar, as drawn in Figure 5a, but is curved. A view of a calculated Fermi surface map over a very wide in-plane momentum range is shown in Figure 5b, where the circular contours of constant $k_z$ centred on ($k_x$=0, $k_y$=0) are also plotted (labelled by their respective multiples of $c^*$) and two examples of how this map changes with excitation energy are shown in Fig. S5 of the Supplementary information. Remarkably, we see that the local VB maxima are sometimes situated in the centre of the pseudo-Brillouin zone at Γ' (for instance, this is the case for the ($k_x$=0, $k_y$=0) VBM which is at the centre of the hexagonal quasi-unit cell indicated on Figure 5b) and sometimes they lie either side of the Γ' point. This reveal again the significant degree of dispersion of the VB in the $c^*$ direction normal to the layers. The surfaces of constant energy are not simple cylinders oriented along $c^*$, but bifurcate periodically as shown in Figure 5a, so that transverse sections through them will show one or two maxima depending on the height of the section in the $c^*$ direction and, thus, the choice of excitation phonon energy.

To visualise this bifurcation better, we have also plotted constant energy surfaces at about 0.2 eV below the VBM for the full 3D Brillouin zone (the red surface in Figure 5a). Where the constant energy surfaces split into two, constant energy surfaces closer to the Fermi level show that there are two global VB maxima located close to the plane of $c^*$ and $b^*_{2D}$ in the volume of the BZ, but not at its surface, and not at any special $k$-points; they are centred in the lobes of the surface shown in Fig. 5a. These maxima are missed in previous calculations of the band structure which have usually focussed on the dispersion along paths between high-symmetry points in the BZ[24]. In Fig. S6 of the Supplementary information, we show calculated valence and conduction band dispersions along the path in the Brillouin zone passing through these two maxima, to confirm that the gap at this point is indirect.

**Discussion**
In studies of the rhenium chalcogenides, much attention has focused on the question of whether ReSe$_2$ and ReS$_2$ possess indirect or direct bandgaps in bulk and monolayer forms. This discussion was based initially on optical studies of few-micron sized bulk samples and was extended to the monolayers as these became available; a consensus is gradually emerging that ReSe$_2$ has an indirect band gap with a valence band maximum located away from the Brillouin zone centre, and that it remains indirect down to one monolayer, whilst ReS$_2$ was claimed until recently to have a direct gap at all thicknesses. First-principles calculations at various levels of approximation have been used to support the experimental studies and it is clear that the positions of the band extrema in such calculations are sensitive to the details of the calculation (in particular, whether or not SOC is included, and what rhenium valence is assumed in DFT) so that previous reports are not entirely consistent.

The present work has tested these ideas and we find that the valence band maxima are indeed located away from the zone centre, as suspected, but that they do not sit precisely on a high symmetry direction and so are easily missed in calculations following conventional paths around special *k* points in the Brillouin zone[24, 45]. Nevertheless, the dispersion in the directions analogous to the two-dimensional hexagonal M and K points is of importance because it shows directly the anisotropy found in experimental studies of optical and transport properties. The key directions are $b_{2D}^*$ (which lies in the real space layer plane, is perpendicular to the Re chains and is a vector in direction Γ–M$_1$) and the vector in direction Γ–K$_2$ which also lies in the plane and is normal to $b_{2D}^*$. Figure 2 showed already that these two directions will display the basic crystal anisotropy very clearly. The measured and calculated valence band dispersions for bulk ReSe$_2$ in these two directions are shown in Figure 4; the colour map in the calculated results indicates the projection of the state onto the Re *d* orbitals, which are the major constituent of the valence band edge. It is clear that the valence band maximum lies off-centre and that there is good quantitative agreement between the band structures in both Γ–M$_1$ and Γ–K$_2$ directions down to a binding energy *E-E$_F$* of at least -2 eV. We emphasize that no fitting has been carried out here; the momentum scale for the simulations is adjusted only by the ratio between the experimental and calculated lattice parameters, so as to scale the Brillouin zones to the same size. This agreement gives confidence in the calculations over the whole Brillouin zone summarised in Figs. 3, 4 and 5. In the Supplementary information, Fig. S7, we show predictions for a ReSe2 monolayer based on the same level of approximation; we find it also to be a highly anisotropic material with an indirect gap; the VBM is located either side of Γ, as the projection of Fig. 5a onto the $k_x$-$k_y$ plane would suggest, and the conduction band minimum is located at Γ.

**Methods**
**ARPES experiments**. Photoemission studies were carried out using the ARPES *k*-microscope of the ANTARES beamline of the SOLEIL synchrotron, equipped with two Fresnel zone plates for focusing of the synchrotron radiation to a beam size of ~100 nm and an order selection aperture to eliminate higher diffraction orders. The nanoscale resolution ensured that monocrystalline regions were probed. The sample was mounted on a nano-positioning stage which allowed both angle-resolved and mapping measurements (the latter were used to locate clean, homogeneous single-crystal regions of the sample). Experiments were performed at photon energies from 95 to 180 eV. ARPES measurements were carried out at room temperature in a vacuum of better than 10$^{-10}$ mbar on a surface cleaved under vacuum.

**Density functional theory calculations.** DFT simulations used the Quantum Espresso[46] (QE) suite of plane-wave codes for total energy and band structure calculations and for post-processing to obtain electronic wavefunctions projected onto atomic bases. XCrysden[47] was used for real and reciprocal space visualisation, including the generation of Figure 1. Fully relativistic pseudopotentials and projector augmented wave (PAW) datasets were generated using QE and PSLibrary[48] for both PBESOL (generalized gradient approximation, GGA) and PZ

(local density approximation, LDA) exchange-correlation functionals; the valence of Re was taken as 15 ($5s^2\ 5p^6\ 5d^5\ 6s^2$). Atomic coordinates were taken from Lamfers *et al.*[8] and were relaxed to obtain forces less than $10^{-3}$ eV Å$^{-1}$. Kinetic energy cutoffs were typically 60 Ry (816 eV) and Monkhorst-Pack[49] *k*-point meshes of 6×6×6 were used; meshes up to 10×10×10 produced no significant changes in the band structures obtained. Results obtained using LDA and GGA are qualitatively similar; for instance, the VB anisotropy and the bifurcation of the VBM appear in both and the main difference, as expected, is in the size of the band gap.

**Sample characterisation**. Samples were obtained from hqgraphene.com and secondary ion mass spectrometry was used by the manufacturers to confirm 99.9995% purity with respect to common impurities including the halogen transport agents used in crystal growth. Samples were studied extensively by Raman spectroscopy[24], confirming their 1T' phase and good crystal quality.

Data supporting this study are available from the University of Bath data archive (DOI: 10.15125/BATH-00332)

**References**

1. Wilson JA, Yoffe AD. The transition metal dichalcogenides: discussion and interpretation of the observed optical, electrical and structural properties. *Adv Phys* **18**, 193-335 (1969).

2. Rahman M, Davey K, Qiao SZ. Advent of 2D Rhenium Disulfide (ReS2): Fundamentals to Applications. *Adv Funct Mater*, (2017).

3. Geim AK, Grigorieva IV. Van der Waals heterostructures. *Nature* **499**, 419-425 (2013).

4. Wang X, Xia F. Van der Waals heterostructures: Stacked 2D materials shed light. *Nat Mater* **14**, 264-265 (2015).

5. Zhu ZY, Cheng YC, Schwingenschlögl U. Giant spin-orbit-induced spin splitting in two-dimensional transition-metal dichalcogenide semiconductors. *Phys Rev B* **84**, 153402 (2011).

6. He K, *et al.* Tightly Bound Excitons in Monolayer WSe2. *Phys Rev Lett* **113**, 026803 (2014).

7. Wang G, *et al.* Spin-orbit engineering in transition metal dichalcogenide alloy monolayers. *Nat Commun* **6**, (2015).

8. Lamfers HJ, Meetsma A, Wiegers GA, deBoer JL. The crystal structure of some rhenium and technetium dichalcogenides. *J Alloys Compd* **241**, 34-39 (1996).

9. Wildervanck JC, Wildervanck F, Jellinek. The dichalcogenides of technetium and rhenium. *J Less-Common Met* **24**, 73-81 (1971).

10. Murray HH, Kelty SP, Chianelli RR, Day CS. STRUCTURE OF RHENIUM DISULFIDE. *Inorganic Chemistry* **33**, 4418-4420 (1994).

11. Alcock NW, Kjekshus A. Crystal structure of ReSe$_2$. *Acta Chem Scand* **19**, 79 (1965).

12. Zhong H-X, Gao S, Shi J-J, Yang L. Quasiparticle band gaps, excitonic effects, and anisotropic optical properties of the monolayer distorted 1T diamond-chain structures ReS2 and ReSe2. *Phys Rev B* **92**, (2015).

13. Ho CH, Huang YS, Chen JL, Dann TE, Tiong KK. Electronic structure of ReS$_2$ and ReSe$_2$ from first-principles calculations, photoelectron spectroscopy, and electrolyte electroreflectance. *Phys Rev B* **60**, 15766-15771 (1999).

14. Ho CH, Huang YS, Liao PC, Tiong KK. Piezoreflectance study of band-edge excitons of ReS2-xSex single crystals. *Phys Rev B* **58**, 12575-12578 (1998).

15. Zibouche N, Philipsen P, Kuc A, Heine T. Transition-metal dichalcogenide bilayers: Switching materials for spintronic and valleytronic applications. *Phys Rev B* **90**, 125440 (2014).

16. Nemilentsau A, Low T, Hanson G. Anisotropic 2D Materials for Tunable Hyperbolic Plasmonics. *Phys Rev Lett* **116**, (2016).

17. Yang SX, Tongay S, Yue Q, Li YT, Li B, Lu FY. High-Performance Few-layer Mo-doped ReSe2 Nanosheet Photodetectors. *Sci Rep* **4**, 6 (2014).

18. Zhong H-X, Gao S, Shi J-J, Yang L. Quasiparticle band gaps, excitonic effects, and anisotropic optical properties of the monolayer distorted 1 T diamond-chain structures ReS 2 and ReSe 2. *Phys Rev B* **92**, 115438 (2015).

19. Yang S, Tongay S, Yue Q, Li Y, Li B, Lu F. High-Performance Few-layer Mo-doped ReSe2 Nanosheet Photodetectors. *Sci Rep* **4**, (2014).

20. Liu E, *et al.* High Responsivity Phototransistors Based on Few-Layer ReS2 for Weak Signal Detection. *Adv Funct Mater* **26**, 1938-1944 (2016).

21. Joshua OI, Gary AS, Herre SJvdZ, Andres C-G. Environmental instability of few-layer black phosphorus. *2D Materials* **2**, 011002 (2015).

22. Lorchat E, Froehlicher G, Berciaud S. Splitting of interlayer shear modes and photon energy dependent anisotropic raman response in n-layer ReSe2 and ReS2. *ACS Nano* **10**, 2752-2760 (2016).

23. Tongay S, *et al.* Monolayer behaviour in bulk ReS2 due to electronic and vibrational decoupling. *Nature Communications* **5**, (2014).

24. Wolverson D, Crampin S, Kazemi AS, Ilie A, Bending SJ. Raman Spectra of Monolayer, Few-Layer, and Bulk ReSe2: An Anisotropic Layered Semiconductor. *Acs Nano* **8**, 11154-11164 (2014).

25. Huang YS, Ho CH, Liao PC, Tiong KK. Temperature dependent study of the band edge excitons of ReS2 and ReSe2. *J Alloys Compd* **262**, 92-96 (1997).

26. Ho CH, Liao PC, Huang YS, Tiong KK. Piezoreflectance study of the band-edge excitons of ReS2. *Solid State Commun* **103**, 19-23 (1997).

27. Ho CH, Liao PC, Huang YS, Tiong KK. Temperature dependence of energies and broadening parameters of the band-edge excitons of $ReS_2$ and $ReSe_2$. *Phys Rev B* **55**, 15608-15613 (1997).

28. Ho CH, Liao PC, Huang YS, Yang TR, Tiong KK. Optical absorption of $ReS_2$ and $ReSe_2$ single crystals. *J Appl Phys* **81**, 6380-6383 (1997).

29. Liang CH, Tiong KK, Huang YS, Dumcenco D, Ho CH. In-plane anisotropic electrical and optical properties of gold-doped rhenium disulphide. *Journal of Materials Science-Materials in Electronics* **20**, 476-479 (2009).

30. Tiong KK, Ho CH, Huang YS. The electrical transport properties of ReS2 and ReSe2 layered crystals. *Solid State Commun* **111**, 635-640 (1999).

31. Yang SX*, et al.* Layer-dependent electrical and optoelectronic responses of ReSe2 nanosheet transistors. *Nanoscale* **6**, 7226-7231 (2014).

32. Ho CH, Hsieh MH, Wu CC, Huang YS, Tiong KK. Dichroic optical and electrical properties of rhenium dichalcogenides layer compounds. *J Alloys Compd* **442**, 245-248 (2007).

33. Jariwala B*, et al.* Synthesis and Characterization of ReS2 and ReSe2 Layered Chalcogenide Single Crystals. *Chem Mater* **28**, 3352-3359 (2016).

34. Liu E*, et al.* Integrated digital inverters based on two-dimensional anisotropic ReS2 field-effect transistors. *Nat Commun* **6**, (2015).

35. Huefner S. *Photoelectron Spectroscopy*. Springer (1996).

36. Fang CM, Wiegers GA, Haas C, deGroot RA. Electronic structures of $ReS_2$, $ReSe_2$ and $TcS_2$ in the real and the hypothetical undistorted structures. *J Phys: Condens Matter* **9**, 4411-4424 (1997).

37. Dumcenco D, Huang YS, Liang CH, Tiong KK. Optical characterization of Au-doped rhenium diselenide single crystals. *Journal of Applied Physics* **104**, (2008).

38. Yu-Ci J, Der-Yuh L, Jenq-Shinn W, Ying-Sheng H. Optical and Electrical Properties of Au- and Ag-Doped ReSe 2. *Japanese Journal of Applied Physics* **52**, 04CH06 (2013).

39. Wang X*, et al.* Enhanced rectification, transport property and photocurrent generation of multilayer ReSe2/MoS2 p–n heterojunctions. *Nano Research* **9**, 507-516 (2016).

40. Zhang E, *et al.* Tunable Ambipolar Polarization-Sensitive Photodetectors Based on High-Anisotropy ReSe2 Nanosheets. *ACS Nano*, (2016).

41. Price P. Two-dimensional electron transport in semiconductor layers. I. Phonon scattering. *Annals of Physics* **133**, 217-239 (1981).

42. Xi J, Long M, Tang L, Wang D, Shuai Z. First-principles prediction of charge mobility in carbon and organic nanomaterials. *Nanoscale* **4**, 4348-4369 (2012).

43. Damascelli A. Probing the electronic structure of complex systems by ARPES. *Phys Scr* **2004**, 61 (2004).

44. Riley JM, *et al.* Direct observation of spin-polarized bulk bands in an inversion-symmetric semiconductor. *Nature Physics* **10**, 835-839 (2014).

45. Zhao H, *et al.* Interlayer Interactions in Anisotropic Atomically-thin Rhenium Diselenide. *arXiv preprint arXiv:150407664*, (2015).

46. Giannozzi P, *et al.* QUANTUM ESPRESSO: a modular and open-source software project for quantum simulations of materials. *J Phys: Condens Matter* **21**, 395502 (2009).

47. Kokalj A. Computer graphics and graphical user interfaces as tools in simulations of matter at the atomic scale. *Comput Mater Sci* **28**, 155-168 (2003).

48. Dal Corso A. Pseudopotentials periodic table: From H to Pu. *Comput Mater Sci* **95**, 337-350 (2014).

49. Monkhorst HJ, Pack JD. Special points for Brillouin-zone integrations. *Phys Rev B* **13**, 5188-5192 (1976).

**Acknowledgements**
Work was supported by the Engineering and Physical Sciences Research Council of the UK under grant EP/P004830/1 and LSH received a studentship under grant EP/L014544/1. JLW was supported under grant EP/M022188. ARPES measurements were provided at the ANTARES beamline of the SOLEIL synchrotron under project 20151237. Computational work was performed on the University of Bath's High Performance Computing Facility.

**Author contributions**
LSH and JLW prepared and characterised the samples and LSH, JLW, CC and JA carried out the ARPES measurements. DW, MCA and JA conceived the measurements and DW carried out the computational work. MMK created Figure 4 and all authors contributed to the interpretation of the data. All authors read and approved the manuscript.

**Additional information**
**Supplementary information** accompanies this paper
Data supporting this study are available from the University of Bath data archive (DOI: 10.15125/BATH-00332

**Competing financial interests.** The authors declare no competing financial interests.